# The Isotopic Field-Charge Assumption Applied to the Electromagnetic Interaction

## György Darvas

Symmetrion, Budapest, H-1067, 29 Eötvös St.
http://symmetry.hu/symmetrion/, darvasg@iif.hu

*Motto*:
"I dreamed that I used to be two cats, and played with each other."
(Frigyes Karinthy, Hungarian writer and humorist, 1887-1938)

**Abstract** This paper applies the isotopic field-charge spin theory (Darvas, IJTP 2011) to the electromagnetic interaction. First there is derived a modified Dirac equation in the presence of a velocity dependent gauge field and isotopic field charges (namely Coulomb and Lorentz type electric charges, as well as gravitational and inertial masses). This equation is compared with the classical Dirac equation. There is shown that, since the presence of isotopic field-charges would distort the Lorentz invariance of the equation, there is a transformation, which together with the Lorenz transformation restores the invariance of the equation, in accordance with the conservation of the isotopic field-charge spin (Darvas, 2009). The paper discusses the conclusions derived from the extensions of the Dirac equation. It is shown that in semi-classical approximation the model provides the original Dirac equation, and at significantly relativistic velocities it approaches to the Schrödinger equation. Among other conclusions, the clue gives physical meaning to the electric moment. The closing section summarises a few further conclusions and shows a few developments to be discussed in detail in a next paper (Darvas, 2013).

## 1 Introduction

The paper starts with discussing the role of the isotopic electric charges in classical EM (sec. 2), then the same in QED (sec. 3), finally, we introduce a kinetic field (**D**, in sec. 4). We discuss the results (the magnetic and electric moments) in a semi-classical case, finally the contribution of the kinetic field to the magnetic and electric moments (sec. 5).

[12,14] developed a general field-theoretical model for the conservation of the isotopic field-charge spin, applicable to different kinds of interaction. It is based on the qualitative distinction between two types of the source charges (e.g., gravitational and inertial masses, etc.) of any physical field, and if so, it assumes interaction between the isotopes of the two types of field-charges (cf., Motto). In 2011, the author started to *specify the model to individual interaction types*. Certain elements were predicted in a phenomenological way earlier [10,11].

[15] presented the isotopic field-charge spin theory and its possible applications first to the description of *gravitational interaction* (at FERT 2011), and it was followed by other papers [13, 17, 18, 19]. It was shown that the presence of a kinetic field, with a velocity dependent metric and isotopic field-charges (namely in that case: distinguished gravitational and inertial masses) require a (velocity arrowed) direction-dependent, anisotropic (that means, Finsler) geometry [16, 17, 19, 21]. Gravity [45] as a Finslerian phenomenon has been discussed recently by other authors as well (e.g., [3, 32]; see also 't Hooft [46] appendices A and B, although there without mentioning Finsler).

Now, the present paper makes *three steps to a further extension of the field theoretic model of the electromagnetic interaction[1]*. First, it introduces the isotopic electric charges in the classical Maxwell EM theory. It shows that applied alone, the presence of isotopic electric charges destroys the Lorentz invariance of the Maxwell theory. To restore the broken invariance, the paper refers to the conservation of the isotopic field-charge spin, proven in [14], so that the two transformations applied together ensure the invariance, and save the physical relevance of the theory. Secondly, the author introduces the isotopic electric charges in the classical Dirac equation [20]. At third, he extends the Dirac equation with a kinetic field, introduced in [12]. The main part of the paper discusses this extended equation. The discussion of the Dirac theory was treated also in [24] by de Haas, E.P.J., nevertheless, there without this kinetic extension. [33] studied also the metric of Dirac's field theory with kinematic conditions, in a similar, but also in a little different context.

---

[1] *Electromagnetic field theories* were related with anisotropic geometries in, at least, two terms. *First*, the Dirac matrices, introduced in QED [27] (1928) follow the rules of hypercomplex numbers (as shown, among others, by [1] Achiezer and Berestetskii, p. 90). Something similar has been introduced by the help of Clifford algebras in [35]. Later Dirac published two essential extensions to his QED theory [28, 29, 30, 31]. At *second*, he introduced a curvilinear co-ordinate system in [31]. So, he defined an auxiliary co-ordinate system $y^A$, "which is kept fixed during the variation process and use the functions $y^i(x)$ to describe the $x$ co-ordinate system in terms of the $y$ co-ordinate system." He defines the $y$ system "so that the metric for the $x$ system" be $g_{\mu\nu} = y_{\Lambda,\mu} y^\Lambda_{,\nu}$. This metric, which then appears in the Hamiltonian of the electromagnetic interaction, was the first step to a later Finsler extension. Some consequences have been discussed in [8], [47], [44] and [40]. Concerning the parallel presence of a scalar and a kinetic gauge field, authors of [42] enlarge the configuration space by including a scalar field additionally, and taking anisotropic models into account too.

If one observes certain formal similarity to the introduction of the $A_\mu$ and $B_\mu$ gauge fields in [49], as well as to the discussion of the proportionality of the $g$ and $g'$ coupling constants to the electron charge – that is not by chance. Nevertheless, the similarity is merely formal. I would like to underline here the differences. The $D_\mu$ gauge field in this paper differs from Weinberg's $B_\mu$ gauge field in its physical interpretation (cf. the derivation of the components of $D_\mu$ in sec. 5.3.2). Concerning the proportionality between the two kinds of coupling constants/electric *charges*, in contrast to [49] this paper places the emphasis on the difference between the two kinds of *masses*. Otherwise, these minor alterations do not influence the conslusions (cf., e.g., (20a) in sec. 7 of [22]).

## 2 Isotopic electric charges in classical EM

According to the isotopic field-charge theory [14], we can replace the charges appearing in our equations by two different (isotopic) charges, a Coulomb-type one, and a kinetic-type one. The Coulomb-type charge is associated with the potential part of the Hamiltonian ($V$), and the kinetic type with the kinetic part of the Hamiltonian ($T$), and they appear in the components of a current density respectively.

In classical electrodynamics, the $A_\mu$ four-potentials of the electromagnetic field were invariant under Lorentz transformation, and the four-current $j_\nu$ components transformed like a vector. We assumed, that the sources of the Coulomb force ($q_V$) are different type charges, than moving charges as sources of currents ($q_T$). The same charges play both roles (cf. covariance), we assume only, that in the two situations they behave as two isotopic states of the same physical property (i.e., field-charge). Provided, that the fourth component (in + + + − signature) of a $j_\nu$ current density, namely $j_4 = ic\rho$ contains a different kind of charge-density ($\rho_V$), than those moving (current-like, kinetic) charges ($\rho_T$) in $j_i$ ($i = 1, 2, 3$), the **j** current would lose its invariance under Lorentz transformation.

We can demonstrate this through the transformation of the Lorentz force:

$$F^\mu = \frac{1}{c} F^{\mu\nu} j_\nu \qquad \text{(where } j_\nu = \rho u_\nu, \text{ and } u_\nu = \frac{dx_\nu}{d\tau} \text{).}$$

In the traditional picture this formulation applies the identical charge density for all components, $F^\mu$ and $j_\nu$ behave like vector components.

Provided, that current-like charges are associated with $\rho_T$ charges, and the real- or Coulomb-charges are $\rho_V$-denoted charges, we should apply $\quad j_i = \rho_T u_i \quad (i = 1, 2, 3)$, and $j_4 = ic\rho_V$ in the proposed isotopic electric charge picture. This latter $j_\nu$ does not transform like a vector, and the electromagnetic force should be written as:

$$F^\mu = F^{\mu\nu} \frac{1}{c} j_\nu = \begin{bmatrix} 0 & B_3 & -B_2 & -iE_1 \\ -B_3 & 0 & B_1 & -iE_2 \\ B_2 & -B_1 & 0 & -iE_3 \\ iE_1 & iE_2 & iE_3 & 0 \end{bmatrix} \begin{bmatrix} \rho_T \dfrac{u_1}{c} \\ \rho_T \dfrac{u_2}{c} \\ \rho_T \dfrac{u_3}{c} \\ i\rho_V \end{bmatrix} = \begin{bmatrix} B_3 \rho_T \dfrac{u_2}{c} - B_2 \rho_T \dfrac{u_3}{c} + E_1 \rho_V \\ -B_3 \rho_T \dfrac{u_1}{c} + B_1 \rho_T \dfrac{u_3}{c} + E_2 \rho_V \\ B_2 \rho_T \dfrac{u_1}{c} - B_1 \rho_T \dfrac{u_2}{c} + E_3 \rho_V \\ iE_1 \rho_T \dfrac{u_1}{c} + iE_2 \rho_T \dfrac{u_2}{c} + iE_3 \rho_T \dfrac{u_3}{c} \end{bmatrix} =$$

$$= \frac{1}{c} \begin{bmatrix} B_3 u_2 - B_2 u_3 & cE_1 \\ -B_3 u_1 + B_1 u_3 & cE_2 \\ B_2 u_1 - B_1 u_2 & cE_3 \\ iE_1 u_1 + iE_2 u_2 + iE_3 u_3 & 0 \end{bmatrix} \begin{bmatrix} \rho_T \\ \rho_V \end{bmatrix} = \frac{1}{c} H^{\kappa l} \rho_l$$

(1)

where $(\kappa = 1, .., 4)$, $(l = 1, 2)$, $E_i = -\partial_i \varphi - \dfrac{1}{c} \dfrac{\partial A_i}{\partial t}$ and $B_i = \mathrm{rot}_i \mathbf{A}$. It is easy to recognise, that the first column of the matrix $H^{\kappa l}$ represents components of the kinetic Lorentz force, wile the second column of the matrix represents components of the Coulomb force.[2] Here $B_i$ are

---

[2] This expression is very close to the approach applied by Dirac [29]. The roots are, however, much elder. To see the origins, I must mention a few other historical steps.

Following the *Symmetry Festival 2003*, when I first discussed the basic ideas – developed in detail in my coming book [16] – with Yuval Ne'eman, there appeared a few similar approach publications. Starting from the fundamental equation by Dirac [29]) $\partial_\mu J^\mu A^\nu = 0$, de Haas, E.P.J. in his PIRT paper [23], for example, derives similar (but not the same) conclusions like we, for QED and the SM, according to which physical real quantities can be derived by the distinction of the (spatially localised) electric potential and the Dirac velocity field. Although, in contrary to Dirac, our theory does not need to assume an ether, we can refer to Dirac's statement [30] where he defines the velocity field through the electromagnetic four-potential: "We have now the velocity at all points of space-time, playing a fundamental part in electrodynamics. It is natural to regard this as the velocity of some real physical thing." While Dirac identifies this "real physical thing" with an ether, our work is an attempt to identify these 'things' with the quanta of a gauge-field, 'localised' in that velocity field. For I received objections since I first communicated the essence of the theory presented later in detail in [14], which objections stated that the assumption of a velocity dependent gauge contradicts localisation, I advise to keep in mind the cited words by Dirac (in addition to my main argument, namely the original formulation of Noether's second theorem [39]). De Haas assumes an analogy between Mie's [37] non-gauge invariant stress-energy tensor, and the stress-energy tensor in Dirac's 1951 theory in a four-velocity field. The analogy works only partially (in my opinion), but the acknowledgement of the role of the velocity field in defining the stress-energy tensor is worth attention, it partially confirms my approach, and leads to the same derivation of the Lorentz transformation of the electromagnetic field components, as I have interpreted it. As de Haas [25] refers to it, the stress-energy tensor by M. von Laue [48] can be written as $T_{\mu\nu} = J_\mu A_\nu$, where

$$A_\nu = \begin{bmatrix} \mathbf{A} \\ \dfrac{i}{c}\Phi \end{bmatrix} \quad \text{and} \quad J_\mu = \begin{bmatrix} \mathbf{J} \\ ic\rho \end{bmatrix}$$

associated only with $\rho_T$, while $E_i$ both with $\rho_V$ and $\rho_T$, where $\varphi = \int \frac{\rho_V}{r} dV$ and $A = \int \frac{j(\rho_T)}{r} dV$ are the retarded scalar and vector potentials. It is obvious from the above matrix equation that this $j_\nu$ is *not* a four-vector, and for $\rho_V$ and $\rho_T$ are mixed during the multiplication by $F^{\mu\nu}$, the components $F^\mu$ do not transform as vector components either.

The result of this example is not in compliance with our experience! With the introduction of the isotopic electric charges, we lost certain symmetry. As a consequence, to restore Lorentz invariance and compliance with experience, our program must include the requirement of the existence of an additional transformation that should counteract the loss of symmetry caused by the introduction of two isotopic states of the charges. This additional transformation was presented in [14, 16] by proving the conservation of the isotopic field-charge spin. The *Lorentz transformation* and the *isotopic field-charge spin transformatio*n combined restore the covariance of the theory. In the next sections, I specify it to the electromagnetic field.

## 3 Isotopic electric charges in QED

I presented the effect of the introduction of isotopic electric charges in the classical electrodynamics in the previous section. Now, as an example, I introduce isotopic field charges in the derivation of the (1928) Dirac equation [27].

Dirac considered in first (unperturbed) approximation a case of no field, when the wave equation reduces to

$$(-p_4^2 + p_i^2 + m^2 c^2)\psi = 0 \qquad (1a)$$

where $p_4 = \dfrac{\mathbf{W}}{c} = ih\dfrac{\partial}{c\partial t}$ and $p_i = -ih\dfrac{\partial}{\partial x_i}$ ($i = 1, 2, 3$) and the wave equation be in the form $(\mathbf{H}-\mathbf{W})\psi = 0$.

---

so

$$T_{\mu\nu} = \begin{bmatrix} \mathbf{J} \\ ic\rho \end{bmatrix} \begin{bmatrix} \mathbf{A} \\ \dfrac{i}{c}\Phi \end{bmatrix} = \begin{bmatrix} \mathbf{J} \otimes \mathbf{A} & \dfrac{i}{c}\Phi\mathbf{J} \\ ic\rho\mathbf{A} & -\rho\Phi \end{bmatrix}$$

what demonstrates an analogy with our formula derived in [22] based on the findings in this paper. Note also that the potential (Coulomb) charges behave like corpuscles, while the kinetic (Lorentz type) charges like waves [16]. This complementary double behaviour (formulated first by Bohr in 1927, then discussed in 1937 [5]) became subject of studies again (cf., [41]).

To maintain the required linearity of the Hamiltonian **H** in $p_\mu$ one introduces the dynamical variables $\alpha_i$ and $\beta$ which are independent of $p_\mu$, i.e., that they commute with $t$, $x_i$. Here Dirac considered particles moving in empty space, so that all points in space were equivalent, and one can expect the Hamiltonian not to involve $t$, and $x_i$. It follows that $\alpha_i$ and $\beta$ are independent of $t$, $x_i$, i.e., that they commute with $p_\mu$ ($\mu = 1, 2, 3, 4$), although this latter held only until we did not distinguish gravitational and inertial masses. Dirac introduced his matrices in order to have other dynamical variables besides the co-ordinates and moments of the electron, so that $\alpha_i$ and $\beta$ may be functions of them, and that the relativistic Lorentz invariant wave function depended on these variables. Dirac's wave equation took the form:

$$(p_4 + \alpha_1 p_1 + \alpha_2 p_2 + \alpha_3 p_3 + \beta)\psi = 0 \qquad (2)$$

According to our assumptions, from here on we should modify the clue followed by Dirac. This equation must lead to a condition where we consider that the interacting two charges are carried by particles with masses in two isotopic field-charge (IFC) states, one of them in potential, the other in kinetic state. Since the masses of the carriers appear explicitly in $\beta$, we have to introduce two kinds of $\beta$, corresponding to the two states: $\beta_V$ and $\beta_T$. We have to note that for the sake of relativistic invariance of the four-momentum's square the mass square in the equation (1a) must be equal with the rest mass. We will see, this is – at least numerically – equal with the potential (gravitational) mass: $m_V = m_0$. We make a qualitative distinction between the masses $m_V$ and $m_T$, where the numerical value of the kinetic mass at relativistic velocities is

$$m_T = \frac{m_0}{\sqrt{1 - \dfrac{v^2}{c^2}}}$$

(here $m_0$ is the rest mass of the particle, and $v$ is the velocity of the particle in kinetic state relative to the interacting other particle in potential state). This qualitative distinction will obtain significance later. Thus the equation (2) leads to

$$(-p_4 + \alpha_1 p_1 + \alpha_2 p_2 + \alpha_3 p_3 + \beta_V)(p_4 + \alpha_1 p_1 + \alpha_2 p_2 + \alpha_3 p_3 + \beta_T)\psi = 0 \qquad (3)$$

In order to agree with Eq. (1) in the form $(-p_4^2 + p_i^2 + m_V m_T c^2)\psi = 0$ – considering, in accordance with [14] that a particle in a kinetic state interacts always with another, which is in potential state – we must demand that the coefficients fulfil the conditions:

(d1)  $\alpha_i^2 = 1$

(d2)  $\alpha_i \alpha_j + \alpha_j \alpha_i = 0 \qquad i \neq j$

(d3)  $\beta_V \beta_T = m_V m_T c^2$

(d4)     $\alpha_i p_i \beta_T + \beta_V \alpha_i p_i = 0$

(d5)     $\beta_V p_4 - p_4 \beta_T = 0$

(d1) and (d2) coincide with conditions established by Dirac. The conditions (d3) – (d5) do not follow from Dirac's original clue. We must discuss them separately.

Considering commutations and introducing the notations $\beta_V = \alpha_4 m_1 c$ and $\beta_T = \alpha_4 m_T c$ (and assume that $\alpha_4$ commutes with $m_V$) then from (d4) and (d3) follows:

(d3*)  $\beta_V \beta_T = \alpha_4 m_V c \alpha_4 m_T c = \alpha_4^2 m_V m_T c^2 = m_V m_T c^2 \rightarrow \alpha_4^2 = 1$,

and in accordance with Dirac, we get from (d1), (d3*), and (d2), (d4)

$\alpha_\mu^2 = 1$,  $\alpha_\mu \alpha_\nu + \alpha_\nu \alpha_\mu = 0$  ($\mu \neq \nu$)  $\mu, \nu = 1, 2, 3, 4$.

These coincide with the conditions deduced by Dirac (the only difference is the qualitative replacement of $m^2$ by $m_V m_T$, taking into account the above notice on the relativistic invariance of the four-momentum's square), and they involve that the derived Dirac matrices will not take different forms in our treatment either.

The condition (d5) deserves some attention. Since there appear qualitatively different $\beta_V$ and $\beta_T$ we cannot commute $\beta$ and $p_4$. We get:

$m_V \dfrac{\partial}{\partial t} \psi = \dfrac{\partial}{\partial t}(m_T \psi)$

and from here:

$(m_T - m_V) \dfrac{\partial \psi}{\partial t} = -\left(\dfrac{\partial m_T}{\partial t}\right)\psi$          (4)

If $m_T$ is stationary, then either $\psi$ is stationary too, or $m_V = m_T$, that means, the gravitational and the inertial masses are of equal value in rest. If the value of $m_T$ is changing in time, then $m_V \neq m_T$. The potential (gravitational) mass coincides with the rest mass of the given particle and is stationary. What is changing along with the velocity (by the Lorentz transformation), that is the kinetic (inertial) mass. We see, the equivalence of gravitational and inertial masses is one of the conditions to derive the Dirac equation, and this condition follows from (d1)-(d5), read from the conditions set up for the $\alpha$ and $\beta$ coefficients.

(The condition (d4) makes possible to conclude a similar expression for the gradient of the $i$-th component of the kinetic mass current in the form, what we can neglect in slowly changing systems, especially near to the rest.

$(m_T - m_V) \dfrac{\partial \psi}{\partial x_i} = -\left(\dfrac{\partial m_T}{\partial x_i}\right)\psi$

However, we cannot disregard it for we discuss a relativistic theory. )

Replacing $\alpha_i$ and $\beta$ with appropriate practical multiplets and notations, Dirac introduced the matrices named after him.

In the presence of an arbitrary electromagnetic field with a scalar potential $\Phi = A_4$ and vector potential $\mathbf{A}$, we substitute  $p_4 + (e_T/c)A_4$ for $p_4$, and $p_i + (e_V/c)A_i$ for $p_i$ in the Hamiltonian for no field, where $e_V$ and $e_T$ denote the potential (Coulomb) and kinetic (Lorentz) charges. Note, that according to the assumption introduced in the IFCS theory [16], there appear potential charges in the scalar field potential ($A_4$), which interacts solely with kinetic charges, and *vice versa*, there appear kinetic charges in the vector field potential ($\mathbf{A}$), which interacts solely with potential charges. Similar to $m_T$, $e_T$ takes also three different values according to the spatial directions, like three components of a three-vector. However, I must mention, that $e_T$ transforms with the velocity in a different way than $m_T$. In fact, it is not just the value of the charge of $e_T$, what changes at relativistic velocities, rather the charge density.

Introducing the above deduced conditions in the equation (3), the Dirac matrices, which follow from those conditions, and make the mentioned replacements to consider the effects of an electromagnetic field on our wave equation, we obtain:

$$\left[ -(p_4 + \frac{e_T}{c}A_4) - \gamma_5(\boldsymbol{\sigma}, \mathbf{p} + \frac{e_V}{c}\mathbf{A}) + \gamma_4 m_V c \right] \cdot \left[ (p_4 + \frac{e_T}{c}A_4) - \gamma_5(\boldsymbol{\sigma}, \mathbf{p} + \frac{e_V}{c}\mathbf{A}) + \gamma_4 m_T c \right] \psi = 0$$

(According to the convention, we replaced the $\rho$ matrices applied in Dirac's original (1928) paper with the more widespread $\gamma$ matrices, so that $\rho_1 = -\gamma_5$ and $\rho_3 = \gamma_4$, and also in accordance with the convention, we replace the original $h$ in Dirac's equations with $\hbar$. To get a more easily comparable equation with the original, derived by Dirac – among other algebraic transformations – we make also the following replacement: $m_V m_T c^2 \equiv m_V^2 c^2 + m_V \left( m_T - m_V \right) c^2$. We use during the transformation of the wave equation that the differential operators are ineffective on the stationary $m_V$ and $e_V$.) We derive:

$$\begin{aligned} \{ [ -(p_4 + \frac{e_T}{c}A_4)^2 + (\mathbf{p} + \frac{e_V}{c}\mathbf{A})^2 + m_V^2 c^2 + \\ + \hbar(\boldsymbol{\sigma}, \mathrm{rot}\left( \frac{e_V}{c}\mathbf{A} \right)) + i\hbar\gamma_5(\boldsymbol{\sigma}, \mathrm{grad}\left( \frac{e_T}{c}A_4 \right) + \frac{1}{c}\frac{\partial}{\partial t}\left( \frac{e_V}{c}\mathbf{A} \right)) ] + \\ + \gamma_4 \left[ -(p_4 + \frac{e_T}{c}A_4) + \gamma_5(\boldsymbol{\sigma}, \mathbf{p} + \frac{e_V}{c}\mathbf{A}) + \gamma_4 m_V c \right] (m_T - m_V)c \} \psi = 0 \end{aligned} \qquad (5)$$

The first three terms in the first [ ] bracket coincide with those in the relativistic wave equation for electromagnetic fields derived by Dirac (with the assumption $m_V = m_0$) with the only difference that we made qualitative distinction between the potential (Coulomb) and kinetic (current-, or Lorentz) charges.

The fourth and fifth terms include $\mathrm{rot}\left( \frac{e_V}{c}\mathbf{A} \right) = e_V\,\mathbf{H}$, where $\mathbf{H}$ is the *magnetic vector* of the field; as well as the *electric vector* of the field in a modified form, where the potential and the kinetic charges are taken into account: $\mathrm{grad}\left( \frac{e_T}{c}A_4 \right) + \frac{\partial}{\partial t}\left( \frac{e_V}{c}\mathbf{A} \right) = e'\,\mathbf{E}$, where $e'$ is a quantum mixture of $e_V$ and $e_T$. The charges appear under the derivation, because the value of $e_T$ changes in relativistic covariant fields (for it is a function of its velocity in the given frame, cf., e.g., [1] §22), and we are free to write $e_V$ also under the time derivative, because the derivative operator has no effect on the potential charge $e_V$. More precisely, it is rather the density of $e_T$, wich changes with its velocity. So, in the following, I will replace the isotopic charges $e_T$ with $\rho_T$ and $e_V$ with $\rho_V$ in the formulas.

The expression in the first [ ] bracket in (5) – with the mentioned alteration in the charges – coincides with the quadratic form of the Dirac equation.

Equation (5) differs from Dirac's result essentially in the last, additional term:

$$-\gamma_4\left[-(p_4 + \frac{\rho_T}{c}A_4) + \gamma_5(\boldsymbol{\sigma}, \mathbf{p} + \frac{\rho_V}{c}\mathbf{A}) + \gamma_4 m_V c\right](m_T - m_V)c\ .$$

This expression can be written by inserting the $p_4$ and $\mathbf{p}$ operators in the following:

$$-\gamma_4\left[-(\frac{i\hbar}{c}\frac{\partial}{\partial t} + \frac{\rho_T}{c}A_4) - \gamma_5(\boldsymbol{\sigma}, i\hbar\,\mathrm{grad} - \frac{\rho_V}{c}\mathbf{A}) + \gamma_4 m_V c\right](m_T - m_V)c$$

The components in this term can be regarded as *the additional energy of the interacting two massive, electrically charged particles due to their assumed additional degree of freedom* (arbitrary positions in the IFCS field). They *express the effect of the relativistic mass increase* – difference between the "dressed" and "bare" masses, i.e., the "dress" in itself – *on the electromagnetic field*. The expression in this last square bracket [ ] coincides again with the Dirac wave equation, in its Hamiltonian form.

This last part of the equation gives account on the *cross-interaction of the electromagnetic field and its two isotopic field charges with the two kinds of isotopic masses in QED*. The state function $\psi$ in this equation, unlike in the original Dirac equation, depends not only on the space-time co-ordinates and the spin, but also on a two valued variable that makes distinction between the isotopic field charges.

In rest (when $m_T = m_V$, $\rho_T = \rho_V$), equation (5) coincides with the Dirac equation. However, in relativistic covariant fields the charges of both the gravitational and the electromagnetic fields will differ not only qualitatively, but also in their quantity, and we must take into account the last term. The appearance of this last term brings in the already acquainted (cf., [14], secs. 3 and 3.2) inconvenient, but not unexpected, asymmetry in our theory that should be counteracted by the presumed new symmetry transformation between the states of the isotopic field charges.

The effects of the operators in the two [ ] square brackets in (5) must be equal:

$$\left[-(p_4 + \frac{\rho_T}{c}A_4)^2 + (\mathbf{p} + \frac{\rho_V}{c}\mathbf{A})^2 + m_V^2 c^2 + \hbar(\boldsymbol{\sigma}, \text{rot}\left(\frac{\rho_V}{c}\mathbf{A}\right)) + i\hbar\gamma_5(\boldsymbol{\sigma}, \text{grad}\left(\frac{\rho_T}{c}A_4\right) + \frac{1}{c}\frac{\partial}{\partial t}\left(\frac{\rho_V}{c}\mathbf{A}\right))\right]\psi =$$

$$= \gamma_4\left[(p_4 + \frac{\rho_T}{c}A_4) - \gamma_5(\boldsymbol{\sigma}, \mathbf{p} + \frac{\rho_V}{c}\mathbf{A}) - \gamma_4 m_V c\right](m_T - m_V)c\psi$$

$$(5a)$$

In the case of classical QED, the left side is equal to 0. The right side is 0, if $m_T = m_V$, that means, in a non-relativistic approximation. The effect of the operator in bracket { } on $\psi$ in Eq. (5) will vanish as a result of the operators in the two square [ ] brackets together. If we demand that our mathematical derivations be in agreement with the time-proven Dirac equation, we must require that the effect of the operators in the first and the second square brackets on the wave function $\psi$ be equal to 0 separately, according to the two sides of the Eq. (5a). Thus our equation (5) separates into two equations.

The *first* equation:

$$[-(p_4 + \frac{\rho_T}{c}A_4)^2 + (\mathbf{p} + \frac{\rho_V}{c}\mathbf{A})^2 + m_V^2 c^2 +$$

$$+\hbar(\boldsymbol{\sigma}, \text{rot}\left(\frac{\rho_V}{c}\mathbf{A}\right)) + i\hbar\gamma_5(\boldsymbol{\sigma}, \text{grad}\left(\frac{\rho_T}{c}A_4\right) + \frac{1}{c}\frac{\partial}{\partial t}\left(\frac{\rho_V}{c}\mathbf{A}\right))]\,\psi = 0 \qquad (6)$$

will provide the solutions of the Dirac equation in the presence of potential and kinetic charges in an electromagnetic field. Note, that there appears only the rest mass ($m_V = m_0$) of the particle. This equation differs from the original Dirac equation only in the presence of the two different isotopic electric charges.

The *second* equation:

$$-\gamma_4\left[-(p_4 + \frac{\rho_T}{c}A_4) + \gamma_5(\boldsymbol{\sigma}, \mathbf{p} + \frac{\rho_V}{c}\mathbf{A}) + \gamma_4 m_V c\right](m_T - m_V)c\psi = 0 \qquad (7)$$

holds either in rest when quantitatively $m_T = m_V$, or when the value in the square bracket is 0.

The form of equation (5) guarantees that in boundary conditions our result coincides with the traditional. In a state close to rest, the second part vanishes and we get back to the well known Dirac equation (6). In extreme relativistic situation, when $m_T \gg m_V = m_0$, (we can neglect the first component in (5), and) we get Eq. (7), and can divide the full modified Dirac equation by ($m_T - m_V$). Eq. (7) can be written in a Schrödinger type form of a wave equation. The Dirac expression in the square bracket in Eq. (7) can be transformed in:

$$i\hbar\frac{\partial}{\partial t}\psi = \left[-\rho_T A_4 - \gamma_5(\boldsymbol{\sigma}, i\hbar c\,\mathrm{grad}-\rho_V\mathbf{A}) + \gamma_4 m_V c^2\right]\psi \qquad (8)$$

where $-\rho_T A_4 - \gamma_5(\boldsymbol{\sigma}, i\hbar c\,\mathrm{grad}-\rho_V\mathbf{A}) + \gamma_4 m_V c^2 = \mathbf{H}$ is the Hamiltonian of the system. There appears only the rest energy of the particles. However, due to the difference between $\rho_T$ and $\rho_V$, this equation cannot be linearised in the four charge current components unless the isotopic field-charge spin invariance rotates the two isotopic electric charges of the electric field into each other in an IFCS gauge field (cf. [14]). This equation does not reflect the effect of that gauge field, because the Dirac equation expresses the interaction of the two electrons in the electromagnetic field, more precisely the scalar Coulomb field with the electromagnetic vector field. In this semi-classical approach[3], I have not taken into account the interaction with the IFCS gauge field.[4]

We can construct the Lagrangian of the interacting coupled two-electron system in the fields of each other from this Hamiltonian. Due to the two kinds of charges, this $L$ differs

---

[3] Later, Dirac [29] considered that the classical theories of electromagnetic field are *approximate* and are valid only if the accelerations of the electrons are small. He stated that the earlier problems of QED resulted not in quantization, rather in the incompleteness of the classical theory of electrons, and one must try to improve on it. For this reason, he proposed to replace the application of the Lorentz condition with a gauge theory. He emphasised also the Hamiltonian approach instead of the Lagrangian one. He introduced a function $\lambda$ (which was different from the quantity introduced by Feynman [34]) and got a current $j_\mu = -\lambda(\partial S / \partial x^\mu + A_\mu^*)$ where $S$ was a gauge function attributed to $A$, and $\lambda$ could be chosen to be an arbitrary infinitesimal at one instant of time, while its value at other times was then fixed by the conservation law $\partial j_\mu / \partial x_\mu = 0$. This method resulted in the conclusion that the theory (as expected) involves only the ratio $e/m$, not $e$ and $m$ separately. This [29] theory did not introduce the interaction of the electron with the electromagnetic field as a perturbation, like in the 1929-1932 Dirac-Fermi-Breit theories. The electron of that new theory could not be considered apart from its interaction with the electromagnetic field. As Dirac mentioned: "The theory of the present paper is put forward as a basis for a passage to a quantum theory of electrons. ... one can hope that its correct solution will lead to the quantization of electric charge ..." and "... questions of the interaction of the electron with itself no longer arise." Then, a further model by Dirac [31] p. 64) provided a possible solution for eliminating the runaway motions of the electron.

Dirac's [29] paper was an attempt to exclude approximations by perturbation in either direction. It was in harmony with the aim of Bethe and Fermi [4] to show the equivalence of the perturbations applied by Breit [7, 8] and Møller [38]. In this respect Dirac's models were kin to the present attempt, in which, instead of a perturbation, we acknowledge the asymmetric roles of the interacting charged particles (as it can be read originally in [38]) and apply a gauge theory that has led us to a quantised theory. Certain ideas are borrowed here from [43]. The theory applied in this paper to QED and having been proposed in a general form in [14] eliminates the runaway motions of the electron too, although in an alternative way.

[4] At the end of their paper Bethe and Fermi ([4] p. 306) showed that the formula introduced by Møller holds also when one of the interacting particles is in bound state. They consider also the option that the two interacting particles emit two quanta, but they reject it, because (for symmetry consideration for the momentums of the two quanta) they take into account only identical type quanta to be emitted and absorbed. (Although, the emission of one quantum painted another asymmetry in the picture, in which they aimed at eliminating the asymmetry caused by Møller's scattering matrix.) This conclusion by Bethe and Fermi is a result of their artificial symmetrisation of the potentials, and does not arise in the theory set forth, among others, in this paper. For those, who are interested in the problem of identity and non-identity of particles, as well as symmetrisation, more detailed analyses are recommended in [26, 16].

from the usual form that appears in most textbooks. In principle, one can derive the non-linear charge–four-current from this $L$. The condition of linearization is that the gauge field, in which the charges $\rho_T$ and $\rho_V$ can substitute for each other, become invariant under an arbitrary gauge transformation. I will consider the interaction with a kinetic, concretely, IFCS gauge field in the next section.

## 4 Isotopic electric charges in the presence of a kinetic gauge field

Let's introduce a kinetic gauge field $\mathbf{D}$ similar to the general field-theoretical approach in [14]. As we saw in section 2 that the vierbein $j_\nu$ does not transform like a vector, we cannot expect this property of $\mathbf{D}$ either. This $\mathbf{D}$ kinetic gauge field is associated with the electromagnetic field. Therefore, I extend the Dirac equation, discussed in section 2, with the components of this $\mathbf{D}$ gauge field. For $\mathbf{D}$ is a kinetic field, all the four of its components interact with the potential electric (Coulomb) charge. Thus, I will start from the following, extended form of the equation:

$$\left[ -(p_4 + \frac{\rho_T}{c} A_4 + \frac{\rho_V}{c} D_4) - \gamma_5(\boldsymbol{\sigma}, \mathbf{p} + \frac{\rho_V}{c} \mathbf{A} + \frac{\rho_V}{c} \mathbf{D}) + \gamma_4 m_V c \right] \cdot$$
$$\cdot \left[ (p_4 + \frac{\rho_T}{c} A_4 + \frac{\rho_V}{c} D_4) - \gamma_5(\boldsymbol{\sigma}, \mathbf{p} + \frac{\rho_V}{c} \mathbf{A} + \frac{\rho_V}{c} \mathbf{D}) + \gamma_4 m_T c \right] \psi = 0 \tag{9}$$

here $D_4$ is the fourth component of $\mathbf{D}$, and $\mathbf{D}$ depends on the velocity components $D_\mu = D(\dot{x}^\mu)$. (Concerning the equivalence of this description and the space-time plus velocity dependent description see [14, 2].) Note, that $D_4$, being a component of the kinetic gauge field interacts with the *potential* electric charge in contrast to the $A_4$ scalar potential of the electric field in the first ( ) brackets, and $\mathbf{D}$ in the second ( ) brackets is a three-component, vector-like quantity. Making the multiplication, applying the same transformations like in section 2, and considering that $p_4 = i\frac{\hbar}{c}\frac{\partial}{\partial t}$ and $p_i = -i\hbar\frac{\partial}{\partial x^i}$, as well as commutation of the components, one gets:

$$\{ [ -(p_4 + \frac{\rho_T}{c} A_4 + \frac{\rho_V}{c} D_4)^2 + (\mathbf{p} + \frac{\rho_V}{c} \mathbf{A} + \frac{\rho_V}{c} \mathbf{D})^2 + m_V^2 c^2 ] +$$
$$+ \hbar(\boldsymbol{\sigma}, \text{rot}\left(\frac{\rho_V}{c} \mathbf{A}\right)) + i\hbar\gamma_5(\boldsymbol{\sigma}, \text{grad}\left(\frac{\rho_T}{c} A_4\right) + \frac{1}{c}\frac{\partial}{\partial t}\left(\frac{\rho_V}{c} \mathbf{A}\right)) +$$
$$+ \hbar(\boldsymbol{\sigma}, \text{rot}\left(\frac{\rho_V}{c} \mathbf{D}\right)) + \hbar(\boldsymbol{\sigma}, \frac{\rho_V^2}{c^2}\left[ D_j D_k - D_k D_j \right]) +$$
$$+ i\hbar\gamma_5(\boldsymbol{\sigma}, \text{grad}\left(\frac{\rho_V}{c} D_4\right) + \frac{1}{c}\frac{\partial}{\partial t}\left(\frac{\rho_V}{c} \mathbf{D}\right)) + \gamma_5\frac{\rho_V^2}{c^2}(\boldsymbol{\sigma}, D_4\mathbf{D} - \mathbf{D}D_4) +$$
$$+ \gamma_4\left[ -(p_4 + \frac{\rho_T}{c} A_4 + \frac{\rho_V}{c} D_4) + \gamma_5(\boldsymbol{\sigma}, \mathbf{p} + \frac{\rho_V}{c} \mathbf{A} + \frac{\rho_V}{c} \mathbf{D}) + \gamma_4 m_V c \right](m_T - m_V)c \} \ \psi = 0 \tag{10}$$

*Eq. (10) is the extended Dirac equation, in the presence of isotopic electric charges and a kinetic gauge field **D***. There was considered that both the components of the EM vector potential **A**, and the elements of **D** commute with **σ**, the components of **A** commute with each other, but, for the elements of **D** do not compose a four-vector (in contrast to the components of **A**), we have no reason to assume that the elements of **D** would commute with each other. Thus, in the multiplication in (9) we considered the following equalities:

$$(\boldsymbol{\sigma}, \mathbf{p} + \frac{\rho_V}{c}\mathbf{A} + \frac{\rho_V}{c}\mathbf{D})^2 = (\mathbf{p} + \frac{\rho_V}{c}\mathbf{A} + \frac{\rho_V}{c}\mathbf{D})^2 + \hbar(\boldsymbol{\sigma}, \mathrm{rot}\left(\frac{\rho_V}{c}\mathbf{A}\right)) +$$

$$+ \hbar(\boldsymbol{\sigma}, \mathrm{rot}\left(\frac{\rho_V}{c}\mathbf{D}\right)) + \hbar(\boldsymbol{\sigma}, \frac{\rho_V^2}{c^2}\left(D_j D_k - D_k D_j\right))$$

and

$$\gamma_5(p_4 + \frac{\rho_T}{c}A_4 + \frac{\rho_V}{c}D_4)(\boldsymbol{\sigma}, \mathbf{p} + \frac{\rho_V}{c}\mathbf{A} + \frac{\rho_V}{c}\mathbf{D}) - \gamma_5(\boldsymbol{\sigma}, \mathbf{p} + \frac{\rho_V}{c}\mathbf{A} + \frac{\rho_V}{c}\mathbf{D})(p_4 + \frac{\rho_T}{c}A_4 + \frac{\rho_V}{c}D_4) =$$

$$i\hbar\gamma_5(\boldsymbol{\sigma}, \mathrm{grad}\left(\frac{\rho_T}{c}A_4\right) + \frac{1}{c}\frac{\partial}{\partial t}\left(\frac{\rho_V}{c}\mathbf{A}\right)) +$$

$$+ i\hbar\gamma_5(\boldsymbol{\sigma}, \mathrm{grad}\left(\frac{\rho_V}{c}D_4\right) + \frac{1}{c}\frac{\partial}{\partial t}\left(\frac{\rho_V}{c}\mathbf{D}\right)) + \gamma_5\frac{\rho_T^2}{c^2}(\boldsymbol{\sigma}, D_4\mathbf{D} - \mathbf{D}D_4)$$

## 5 Discussion of the modified Dirac equation in the presence of isotopic electric charges and a kinetic gauge field

Equation (10) can be written in the following form: $[W + W^A + W^D - H(m_T - m_V)c]\psi = 0$, where $W$ refers to the first line of (10), $W^A$ to the second line, $W^D$ to the third and fourth lines, and $H(m_T - m_V)c$ to the fifth line of Eq. (10).

### 5.1 Coincidence with the classical Dirac equation in boundary case, when no kinetic field is present

The first line of the operator in Eq. (10), $W$ expresses the first three elements of the classical Dirac equation, with the modifications that it contains (a) the isotopic electric charges, and (b) the kinetic vector potential **D** of the considered kinetic field.

### 5.2 The magnetic and the electric moments

The two elements in $W^A$ – considering the isotopic electric charges – coincide with the two elements of the classical Dirac equation as discussed in section 3, and yield the *magnetic* and the *electric moments* of the electron interacting with the electromagnetic field, respectively.

### 5.3 The magneto-kinetic and electro-kinetic moments

The essential difference, compared to the Eq. (5) of the semi-classical QED model discussed in section 3, is in $W^D$ expressed in the lines 3 and 4 of the Eq. (10). The expression

$$\hbar(\boldsymbol{\sigma}, \mathrm{rot}\left(\frac{\rho_V}{c}\mathbf{D}\right)) + \hbar(\boldsymbol{\sigma}, \frac{\rho_V^2}{c^2}\left[D_j D_k - D_k D_j\right]) +$$
$$+ i\hbar\gamma_5(\boldsymbol{\sigma}, \mathrm{grad}\left(\frac{\rho_V}{c}D_4\right)) + \frac{1}{c}\frac{\partial}{\partial t}\left(\frac{\rho_V}{c}\mathbf{D}\right)) + \gamma_5 \frac{\rho_V^2}{c^2}(\boldsymbol{\sigma}, D_4\mathbf{D} - \mathbf{D}D_4) \qquad (11)$$

provides a *kinetic moment* of the **D** field. Introducing the commutator of **D**, one can write the following:

$$\hbar(\boldsymbol{\sigma}, \mathrm{rot}\left(\frac{\rho_V}{c}\mathbf{D}\right)) + ig\hbar\frac{\rho_V^2}{c^2}(\boldsymbol{\sigma}, C_{jk}^i D_j D_k) +$$
$$+ i\hbar\gamma_5(\boldsymbol{\sigma}, \mathrm{grad}\left(\frac{\rho_V}{c}D_4\right)) + \frac{1}{c}\frac{\partial}{\partial t}\left(\frac{\rho_V}{c}\mathbf{D}\right)) + \gamma_5 \frac{\rho_V^2}{c^2}(\boldsymbol{\sigma}, D_4\mathbf{D} - \mathbf{D}D_4) \qquad (11a)$$

or

$$(\frac{\hbar}{c}\boldsymbol{\sigma}, \mathrm{rot}\left(\rho_V\mathbf{D}\right)) + (\frac{\hbar}{c}\boldsymbol{\sigma}, igC_{jk}^i \frac{\rho_V^2}{c}D_j D_k) +$$
$$+ i\gamma_5\left[\left(\frac{\hbar}{c}\boldsymbol{\sigma}, \mathrm{grad}\left(\rho_V D_4\right) + \frac{1}{c}\frac{\partial}{\partial t}\left(\rho_V\mathbf{D}\right)\right) + \left(\frac{\hbar}{c}\boldsymbol{\sigma}, \frac{\rho_V^2}{\hbar c}(\mathbf{D}D_4 - D_4\mathbf{D})\right)\right] \qquad (11b)$$

Here $C_{jk}^i$ are the structure constants appearing with the multiplication of $D_i$-s, and $g$ is the coupling constant for the electromagnetic interaction. $C_{jk}^i$ are the coefficients in the commutation rule of the generators (transformation matrices) of the symmetry group of the kinetic (isotopic field-charge) field, as we saw in [14]. Since this field is subject of an SU(2) symmetry, there are three $C_{jk}^i$ structure constants. This commutation term does not appear in $W^A$, because the **A** vector potential of the EM field composes a vector, and the derivatives of **A** commute with each other as vectors.

Since the derivatives of $D_\mu = D(\dot{x}^\mu)$ appearing in (10) are derived by the space-time co-ordinates, and $D_\mu$ depends on $\dot{x}^\mu$, all derivatives of $D_\mu$ must be interpreted as

$$\frac{\partial D_\mu}{\partial x^\nu} = \frac{\partial D_\mu}{\partial \dot{x}^\rho}\frac{\partial \dot{x}^\rho}{\partial x^\nu} = D_{\mu,\rho}\lambda_\nu^\rho = D_{\mu,\rho}\dot{x}_{,\nu}^\rho \quad \text{(where } \mu, \nu, \rho = 1, ..., 4)$$

For simplicity, let us assume that **D** depends only on the linear combinations of the first derivatives and multiplications of the velocity, on the velocity itself, as well as on

$$\kappa = \frac{1}{\sqrt{1 - \frac{\dot{x}_i^2}{c^2}}} \text{ and a constant. So}$$

$$D(\dot{x}^{\mu}) = \alpha \frac{\partial \ddot{x}^{\mu}}{\partial \dot{x}^{\rho}} \frac{\partial \ddot{x}^{\rho}}{\partial \dot{x}^{\nu}} + \beta \dot{x}_i \dot{x}_j + \gamma \dot{x}_i + \delta \kappa + \varepsilon$$

where $\alpha$, $\beta$, $\gamma$, $\delta$ and $\varepsilon$ are coefficients, not depending on the actual relative velocity of the interacting charges. In this plausible, but quite not the most general case, the commutator of $D_j$ and $D_k$ is not identically 0. However, all the three elements of $D_i$ ($i = 1, 2, 3$) commute with $D_4$. In this case the third term in $\mathbf{N}^D$ (see below) vanishes.

In the general case, (11b) can be written as $(\frac{\hbar}{c}\boldsymbol{\sigma}, \mathbf{M}^D) + i\gamma_5(\frac{\hbar}{c}\boldsymbol{\sigma}, \mathbf{N}^D)$, where

$$\mathbf{M}^D = \text{rot } \rho_V \mathbf{D} + igC^i_{jk}\frac{\rho_V^2}{c}D_jD_k \quad \text{and} \quad \mathbf{N}^D = \text{grad } \rho_V D_4 + \frac{1}{c}\frac{\partial}{\partial t}\rho_V \mathbf{D} + \frac{\rho_V^2}{\hbar c}(\mathbf{D}D_4 - D_4\mathbf{D}) \quad (11c)$$

Note, that in the expressions of $\mathbf{M}^D$ and $\mathbf{N}^D$, there appear only the potential (Coulomb) charge densities. This is natural, because the velocity dependence is considered in the kinetic gauge potential $\mathbf{D}$, which these charges interact with. Since $\rho_V$ does not depend either on space-time co-ordinates, or on the actual velocity, it is not subject of derivation:

$$\mathbf{M}^D = \rho_V \text{ rot } \mathbf{D} + igC^i_{jk}\frac{\rho_V^2}{c}D_jD_k \quad \text{and} \quad \mathbf{N}^D = \rho_V \text{ grad } D_4 + \frac{\rho_V}{c}\frac{\partial}{\partial t}\mathbf{D} + \frac{\rho_V^2}{\hbar c}(\mathbf{D}D_4 - D_4\mathbf{D})$$

*The kinetic moment is an additional, new quantity in the isotopic electric charge theory* compared to the classical Dirac theory. *The two kinetic moments determine the isotopic electric charge spin $\Delta_{el}$.* According to [14], the isotopic field-charge spin (including also the isotopic electric charge spin) is a conserved quantity, so it must commute with the Hamiltonian.

### 5.3.1 The magneto-kinetic moment

The first term in Eq. (11), $\mathbf{M}^D$ (with an ($m_T$–$m_V$) divider) can be considered as a "magneto-kinetic" additional energy of the electric charge due to its additional degree of freedom assigned to it by the interaction with the kinetic field. For $D_{ji}$ does not behave as a vector, its derivatives include an additional, gauge term, what the derivation of the extended Dirac equation (10) provided automatically in the form of the third term ($W^D$) in the expression (10). So, this second term of $\mathbf{M}^D$ forms part of the "magneto-kinetic" momentum of the field (cf., the first line of Eq. (11)).

The full magnetic moment of the interaction (with a mass-dimension divider), in the presence of the kinetic gauge field, will be:

$$(\frac{\hbar}{c}\boldsymbol{\sigma}, \mathbf{M}^{FULL}) = (\frac{\hbar}{c}\boldsymbol{\sigma}, \rho_V \text{ rot}(\mathbf{A} + \mathbf{D}) + igC^i_{jk}\frac{\rho_V^2}{c}D_jD_k)$$

### 5.3.2 The electro-kinetic moment

The third term in Eq. (11) is similar to the expression got for the electric moment of the EM field in the line 2 ($W^A$) of (10), extended also with a gauge term. It can be considered in a similar way, like in Dirac's theory, as an "electro-kinetic" additional energy of the electron. Also, similar to the $i\hbar\gamma_5\left(\boldsymbol{\sigma}, \operatorname{grad}\left(\dfrac{\rho_T}{c}A_4\right) + \dfrac{1}{c}\dfrac{\partial}{\partial t}\left(\dfrac{\rho_V}{c}\mathbf{A}\right)\right)$ electric moment in $W^A$, the "electro-kinetic" moment

$$i\gamma_5\left[\left(\frac{\hbar}{c}\boldsymbol{\sigma}, \rho_V \operatorname{grad} D_4 + \frac{\rho_V}{c}\frac{\partial}{\partial t}\mathbf{D}\right) + \left(\frac{\hbar}{c}\boldsymbol{\sigma}, \frac{\rho_V^2}{\hbar c}(\mathbf{D}D_4 - D_4\mathbf{D})\right)\right]$$

is apparently imaginary too. However, this is only an appearance. Dirac observed the following: "It is doubtful whether the electric moment has any physical meaning" as a result that the multiplication, due to which the imaginary term appeared, was an artificial involvement in the equation. Note, that while $\sigma_1$ and $\sigma_3$ are real, $\sigma_2$ is imaginary, so $i\sigma_2$ is real, and only $i\sigma_1$ and $i\sigma_3$ are imaginary. (At the same time, the second component in the magnetic moment is also imaginary, due to the imaginary character of $\sigma_2$.) In contrast to the original Dirac equation in the case of the extended Dirac equation, the multiplier of the three-vector $\boldsymbol{\sigma}$ (or $\gamma_5\boldsymbol{\sigma}$) is a sum, which includes the components of the velocity-dependent field. In the presence of the kinetic field $\mathbf{D}$, one can choose the co-ordinate system fitted to the electron's velocity arrow so that the multiplier of the imaginary $\sigma_2$ be non-zero. Then, the expression will yield *a real term* for the sum of the electric and the electro-kinetic energy, while there will be left *two imaginary terms* for the multiplication by $\sigma_1$ and $\sigma_3$, whose sum should be made equal to 0, and provide a constraint for the energy of the interaction. Note, that these two expressions are not fully imaginary, since $D_4$, depending on the fourth component of the velocity, is imaginary itself. On the other side, the multiplier of $\sigma_2$ contains an imaginary component, too ($\operatorname{grad}_2 D_4$). With these conditions, *one can eliminate the imaginary terms* in (10) and *give physical meaning to the electric and the electro-kinetic moments*.

$$i\hbar\gamma_5\left(\sigma_1, \operatorname{grad}_1\left(\frac{\rho_T}{c}A_4\right) + \frac{1}{c}\frac{\partial}{\partial t}\left(\frac{\rho_V}{c}A_1\right)\right) + i\hbar\gamma_5\left(\sigma_1, \operatorname{grad}_1\left(\frac{\rho_V}{c}D_4\right) + \frac{1}{c}\frac{\partial}{\partial t}\left(\frac{\rho_V}{c}D_1\right)\right) +$$

$$+ \gamma_5\frac{\rho_V^2}{\hbar c^2}\left(\sigma_1, D_4 D_1 - D_1 D_4\right) = 0$$

$$i\hbar\gamma_5\left(\sigma_3, \text{grad}_3\left(\frac{\rho_T}{c}A_4\right) + \frac{1}{c}\frac{\partial}{\partial t}\left(\frac{\rho_V}{c}A_3\right)\right) + i\hbar\gamma_5\left(\sigma_3, \text{grad}_3\left(\frac{\rho_V}{c}D_4\right) + \frac{1}{c}\frac{\partial}{\partial t}\left(\frac{\rho_V}{c}D_3\right)\right) +$$

$$+\gamma_5\frac{\rho_V^2}{\hbar c^2}\left(\sigma_3, D_4 D_3 - D_3 D_4\right) = 0$$

The two conditions of the above equation can be satisfied, if both the imaginary and the real parts are equal to 0 separately.

$$i\hbar\left[\text{grad}_1\left(\rho_T A_4\right) + \frac{\rho_V}{c}\frac{\partial}{\partial t}\left(A_1 + D_1\right)\right] + \frac{\rho_V^2}{\hbar c}\left(D_4 D_1 - D_1 D_4\right) = 0$$

$$\text{grad}_1 D_4 = 0$$

and

$$i\hbar\left[\text{grad}_3\left(\rho_T A_4\right) + \frac{\rho_V}{c}\frac{\partial}{\partial t}\left(A_3 + D_3\right)\right] + \frac{\rho_V^2}{\hbar c}\left(D_4 D_3 - D_3 D_4\right) = 0$$

$$\text{grad}_3 D_4 = 0$$

We add from among the multipliers of $\sigma_2$: $\qquad\qquad \text{grad}_2 D_4 = 0$

Provided that the components of the **A** vector potential and the value of the interacting charges are known, we have got a set of differential equations to determine the components of $D_\mu$, and the actual value of the velocity arrowed parallel to $\sigma_2$, in each space-time point.

The reference frame, in which we calculated the constraints for **D**, rotates together with the kinetic charge density $\rho_T$. This choice provided a restriction for the interacting system, while at the same time, it made us possible to calculate the exact forms of the components of $D_\mu$ in the given reference frame.

Considering also the assumption formulated in section 5.3, the set of differential equations reduces to the following:

$$\text{grad}_1\left(\rho_T A_4\right) + \frac{\rho_V}{c}\frac{\partial}{\partial t}\left(A_1 + D_1\right) = 0; \quad \text{grad}_1 D_4 = 0$$

$$\text{grad}_2 D_4 = 0 \qquad\qquad (11d)$$

$$\text{grad}_3\left(\rho_T A_4\right) + \frac{\rho_V}{c}\frac{\partial}{\partial t}\left(A_3 + D_3\right) = 0; \quad \text{grad}_3 D_4 = 0$$

From the non-zero multiplier of $\sigma_2$ we get

$$\text{grad}_2\left(\frac{\rho_T}{c}A_4\right) + \frac{1}{c}\frac{\partial}{\partial t}\left[\frac{\rho_V}{c}\left(A_2 + D_2\right)\right] \neq 0,$$

which, as we will see, is equal to $\mathbf{N}^{FULL}$.

This set of differential equations, extended with the formula for $\mathbf{M}^D$ yield the four components of $D_\mu$, and the kinetic charge current density $\rho_T$, which latter depends on the actual relative velocity of the two-charge system. In this case the electro-kinetic moment (with a mass-dimension divider) will take the form

$$i\gamma_5(\frac{\hbar}{c}\boldsymbol{\sigma},\mathbf{N}^D) = i\gamma_5\rho_V\left[\frac{\hbar}{c}\boldsymbol{\sigma}, \operatorname{grad} D_4 + \frac{1}{c}\frac{\partial}{\partial t}\mathbf{D} + \frac{\rho_V}{\hbar c}(\mathbf{D}D_4 - D_4\mathbf{D})\right]$$

where the first and third terms in the right side are equal to 0, so the electro-kinetic moment (with a mass-dimension divider) will reduce to:

$$i\gamma_5(\frac{\hbar}{c}\boldsymbol{\sigma},\mathbf{N}^D) = i\gamma_5\rho_V\left(\frac{\hbar}{c}\boldsymbol{\sigma}, \frac{1}{c}\frac{\partial}{\partial t}\mathbf{D}\right)$$

The full electric moment (with an $m_T$–$m_V$) divider) can be written as:

$$i\gamma_5(\frac{\hbar}{c}\boldsymbol{\sigma},\mathbf{N}^{FULL}) = i\gamma_5\left(\frac{\hbar}{c}\boldsymbol{\sigma}, \operatorname{grad}\rho_T A_4 + \frac{\rho_V}{c}\frac{\partial}{\partial t}(\mathbf{A}+\mathbf{D})\right),$$

whose second component ($i\sigma_2$ multiplied) is real, the first and third components are 0. The electric moment of the interacting particles is directed towards the real component of the spin ($i\sigma_2$). We have not experienced such moment in the classical Dirac theory.[5] In contrast to the Pauli-Dirac treatment [27], this moment is a measurable quantity. Further conclusions are discussed in [22].

## 6 Concluding remarks

(1) The potential mass and charge appear in the Coulomb potential and do not change with increasing velocity. The kinetic mass and charge appear in the kinetic part of the Hamiltonian, and change with the velocity. All other terms in the Hamiltonian behave in a similar way. Therefore, the two kinds of constituents of the Hamiltonian cannot be contracted, like in the classical Dirac QED. The difference between the potential and kinetic charges and masses were considered by Weinberg [49] integrated in the two kinds of coupling constants (cf., his Eq. (15), p. 1265).
Weinberg described a renormalisable model of spontaneously broken symmetry between electromagnetic and weak interactions, by a two step perturbation, introducing the photon and the weak intermediate bosons as gauge fields. At the end of his paper he put the question: "... what happens if we extend it to include the couplings of $A_\mu$ and $B_\mu$ to the hadrons?" (p. 1266). That was a first step towards the unification of his combined electroweak theory with the strong interaction. The present paper derived the electromagnetic theory by the help of another two-step perturbation ($A_\mu$ and $D_\mu$), starting from a kinetic extension of the Dirac equation.

(2) The novelties in this paper can be most easily understood by starting with the Eq. (10).
$$[W+W^A+W^D-H(m_T\text{–}m_V)c]\psi=0$$

---

[5] The fact is, that Dirac ([27] p. 619) could not do anything with the electric moment, and so did all but most textbooks following him. The appearance of the kinetic field $\mathbf{D}$ made possible to calculate the full electric moment.

The last term has the form of a Schrödinger equation (cf., Eq. (12), which plays an independent role with conditions when the first three terms can be neglected, cf., 5b below). It has appeared in the Dirac equation never before!

(3) The 3rd term ($W^D$) is a contribution to the Electric moment. (Dirac was mistaken, when he wrote that the Electric moment was imaginary. In fact, it was only the second component of $\sigma$ which was imaginary. But the importance lies not in this.) This additional term to the Electric moment allowed us to determine the real part of the Electric moment. Making the imaginary parts equal to 0, we got a set of differential equations to explicitly identify the form of the unknown quantities (sec. 5.3.2). So, **D** is fully determined.

(4) Dirac was aware that his theory was an approximation. Many of his followers were not. He tried to extend his theory twice (1951 and 1962). He failed both times.
C. Møller (1931) showed the here treated asymmetry between the roles of the interacting isotopic electric charges. Unfortunately, Bethe (and Fermi, 1932) symmetrised it mistakenly.
This asymmetry was used also by S. Weinberg (1967) in his chiral electro-weak unification theory without any reference to C. Møller.

(5a) To justify the approximative character of the „semi-classical" Dirac theory, let us first consider an unperturbed, classical case, when $m_V = m_T$. In this case, the last three lines in Eq. (10) vanish, and we get back to the "semi-classical" Dirac equation. All experimental data were obtained for this slightly relativistic situation. Now, we learned, that was only an approximation when one disregards the additional terms. However, this extended Dirac equation includes, as a semi-classical boundary case, the original equation (cf., line 1 in Eq. (10)).

(5b) In an "extremely" relativistic case, when $m_T \gg m_V$, the first four lines of Eq. (10) can be neglected. In this case, the situation reduces to the 5th line, which is identical with the Schrödinger equation. Its validity has also been confirmed many times.

Now, the situation is similar to that in the time of the first formulation of Planck's quantum hypothesis for the black hole radiation, set up in the Fall of 1900. We have experimental tests for the two ends of a kinetic energy scale. But we couldn't have to fit them at the middle. Now we can!

(5c) Therefore, the most interesting situation is, when $m_T > m_V$, but not "too much". This situation is discussed in Sections 5-8 in detail. Lines 3-4 ($W^D$) in Eq. (10) get importance in this instance. This is the part, which was disregarded in Dirac's ([27] 1928) paper (and in the majority of textbooks, which refer to it, up to now), and this is to which he refers so, that his theory is "only" an approximation (cf., footnote 3). This situation has not been tested, because it has not been discussed. This test should be executed. I give my simplified proposals in Section 11.

(6) Based on the calculation of the components of the **D** field in section 5, we have determined the extended, full magnetic and kinetic moments in an electromagnetic interaction. On that basis, are we able to calculate a new Hamiltonian and Lagrangian of the interaction [22], and the full magneto-kinetic and electro-kinetic moments [22]. Both are real, measurable physical quantities. This restored the physical meaning of the electric moment rejected by Dirac in [27] (cf., footnote 5).